\documentstyle[aps,graphicx]{revtex}
\begin{document}
\draft
\title{Transitions to the Fulde-Ferrell-Larkin-Ovchinnikov phases at low 
temperature in two dimensions}
\author{R. Combescot and C. Mora}
\address{Laboratoire de Physique Statistique,
 Ecole Normale Sup\'erieure*,
24 rue Lhomond, 75231 Paris Cedex 05, France}
\date{Received \today}
\maketitle

\begin{abstract}
We explore the nature of the transition to the Fulde-Ferrell-Larkin-
Ovchinnikov superfluid phases
in the low temperature range in two dimensions, for the simplest 
isotropic BCS model. This is done by applying the Larkin-Ovchinnikov 
approach to this second order transition. We show that there is a 
succession of transitions toward ever more complex order parameters 
when the temperature goes to zero. This gives rise to a cascade with, in 
principle, an infinite number of transitions. Except for one case, the 
order parameter at the transition is a real superposition of cosines with 
equal weights. The directions of these wavevectors are equally spaced 
angularly, with a spacing which goes to zero when the temperature goes 
to zero. This singular behaviour in this $ T = 0$ limit is deeply linked to 
the two-dimensional nature of the problem.
\end{abstract}

\section{INTRODUCTION}

The possible existence of the Fulde-Ferrell-Larkin-Ovchinnikov (FFLO) 
superfluid phases \cite{ff,larkov} has been pointed out in the early 
sixties and it has given rise to much work ever since that time. In 
addition to their intrinsic fundamental interest \cite{cana}, 
these phases are quite 
relevant experimentally since they are expected to arise in 
superconductors with very high critical fields, which are naturally very 
actively searched for. On several occasions these phases have been 
claimed to be observed experimentally, but to date these hopes have not 
been firmly substantiated. Very recently anomalies in the heavy fermion 
compound CeCoIn$_5$ have been attributed to FFLO phases 
\cite{movsh}. The case of two-dimensional (2D) systems is of 
particular interest \cite{shima1} since they are experimentally quite 
relevant. Indeed a major strategy to observe these transitions is to 
eliminate orbital currents, which are responsible for the low critical 
fields in standard superconductors. This can be achieved in quasi two-
dimensional systems, made of widely separated conducting planes, 
such as organic compounds or high $T_{c}$  cuprate superconductors. 
In this case hopping between planes is very severely restricted. Hence 
the orbital currents perpendicular to the planes are very weak when a 
strong magnetic field is applied parallel to the planes, and there is 
essentially no orbital pair breaking effect which opens the path to FFLO 
phases at much higher fields. Indeed experimental results in organic 
compounds have been claimed quite recently \cite{singleton,tanatar} to 
be compatible with the existence of FFLO phases. Naturally when the magnetic
field is not exactly parallel to the planes, one finds in addition
vortex-like structures and the physical situation gets even more complex
\cite{klein}.

The FFLO transition in 2D systems is believed to be second order and 
in particular Burkhardt and Rainer \cite{br} have studied in details the 
transition to a planar phase, where the order parameter $\Delta ( {\bf  
r})$ is a simple $ \cos({\bf  q}. {\bf  r})$ at the transition. This phase 
has been found by Larkin and Ovchinnikov \cite{larkov} to be the best 
one in 3D at $T=0$ for a second order phase transition. And in 3D it is 
also found to be the preferred one in the vicinity of the tricritical point 
and below \cite{matsuo,buz1,cm}, although in this case the transition 
turns out to be first order (except at very low temperature). However it 
is not clear that this is always the case since, as first explored by Larkin 
and Ovchinnikov, this order parameter is in competition with any 
superposition of plane waves, provided that their wavevectors have all 
the same modulus. Indeed we have shown very recently \cite{cm3d} that, 
at low temperature, the transition is rather a first order one,
toward an order parameter 
with a more complex structure. For example at $T=0$ it is very near the 
linear combination of three cosines oscillating in orthogonal directions.

In this paper we explore the low temperature range in 2D and show that 
the second order transition is indeed toward rather more complex order 
parameters. A short report of our results has already been published \cite{mc}.
A first step in this direction is found in the recent work of 
Shimahara \cite{shima2} who found a transition toward a superposition 
of three cosines. Here we show that, when the temperature is lowered 
toward $T=0$, one obtains a cascade of transitions toward order 
parameters with an ever increasing number of plane waves. The $T=0$ 
limit is singular in this respect. This is actually clear from the 
beginning. Indeed if one looks at the second order term in the 
expansion of the free energy in powers of the order parameter, which 
gives the location of the FFLO transition, one finds it to be a singular 
function of the plane wave wavevector. This is recalled in the next 
section. Then we calculate the fourth order term in the free energy 
expansion and show that the phases which are selected by this term 
display the cascade of transitions mentionned above.

\section{THE FREE  ENERGY EXPANSION : SECOND ORDER 
TERM}

The general expression for the free energy difference $\Omega \equiv  
\Omega _{s} - \Omega _{n}$ between the superconducting and the 
normal state can be obtained in a number of ways, starting for example 
\cite{br,eil,werth}  from Eilenberger's expression in terms of the 
quasiclassical Green's function or from the gap equation \cite{larkov} 
and Gorkov's equations. When the result is expanded up to fourth 
order term in powers of the Fourier components $\Delta _{{\bf  q}}$ of 
the order parameter $\Delta ({\bf  r})$: 
\begin{eqnarray}
\Delta ({\bf  r}) = \sum _{{\bf  q}_{i}} \Delta _{{\bf  q}_{i}} 
\exp(i{\bf  q}_{i}. {\bf  r})
\label{eq1}
\end{eqnarray}
one obtains:
\begin{eqnarray}
\frac{\Omega}{ N _{0}} = \sum_{{\bf  q}} \Omega _{2}(q, \bar{ \mu 
},T)
 | \Delta _{{\bf  q}} |^{2}
+   \frac{1}{2} \sum_{{\bf  q}_{i}} \Omega _{4}({{\bf  
q}}_{1},{{\bf  q}}_{2},{{\bf  q}}_{3},{{\bf  q}}_{4},\bar{ \mu 
},T) \Delta _{{\bf  q}_{1}} \Delta ^{*} _{{\bf  q}_{2}} \Delta _{{\bf  
q}_{3}}  \Delta ^{*} _{{\bf  q}_{4}}
\label{freeener}
\end{eqnarray}
where we have momentum conservation ${\bf  q}_{1} + {\bf  q}_{3} 
= {\bf  q}_{2} + {\bf  q}_{4}$ in the fourth order term and $ N _{0}$ 
is the single spin density of states at the Fermi surface. The explicit 
expression of $ \Omega _{2}(q, \bar{ \mu },T) $ in terms of the 
standard BCS interaction  $V$ and of the free fermions propagator is:
\begin{eqnarray}
N _{0} \Omega _{2}(q, \bar{ \mu },T) =  \frac{1}{V} - 
T  \sum_{n,k}  \bar{G}({\bf k}) G({\bf k}+{\bf q})
\label{eq3}
\end{eqnarray}
where $ G( {\bf k}) = (i \bar{\omega  } _{n} - \xi _{{\bf  k}}) ^{-1}$ 
and $ \bar{G}( {\bf k}) = (-i \bar{\omega  } _{n}  - \xi _{{\bf  k}}) 
^{-1}$ and $\bar{\omega  }_{n} \equiv \omega _{n} - i \bar{\mu }$, 
with  $ \bar{ \mu } = (\mu _{\uparrow} - \mu_{\downarrow})/2 $ 
being half the chemical potential difference between the two fermionic 
populations forming pairs, $\xi _{ {\bf  k}}$ the kinetic energy 
measured from the Fermi surface for $ \bar{ \mu } = 0 $ and $ \omega 
_{n} = \pi T (2n+1) $ the Matsubara frequency. Performing the $ \xi 
_{{\bf  k}}$ integration and the 2D angular average over $ \hat{\bf k}$ 
gives:
\begin{eqnarray}
\Omega _{2}(q, \bar{ \mu },T) =   \frac{1}{N _{0}V} + 2 \pi  T  \: 
{\rm Im} 
\sum_{n=0} ^{ \omega _{c}}  \frac{1}
{\sqrt{( i \bar{\omega  } _{n})^{2}-\bar{ \mu }^{2}  \bar{q}^{2}}} 
\label{eq4}
\end{eqnarray}
where we have introduced the dimensionless wavevector $ \bar{q} = q 
v_{F} / 2 \bar{ \mu } $. In Eq.(4) the summation has to be cut-off at a 
frequency $ \omega _{c}$ in the standard BCS way. It is more 
convenient to rewrite $\Omega _{2}$, by introducing physical 
quantities related to the $ \bar{q}= 0 $ case, as:
\begin{eqnarray}
\Omega _{2}(q, \bar{ \mu },T) =  a _{0}(\bar{\mu },T) + I(q, \bar{ 
\mu },T)
\label{eq5}
\end{eqnarray}
with:
\begin{eqnarray}
I(q, \bar{ \mu },T) = 2 \pi  T  \: {\rm Im} 
\sum_{n=0} ^{ \infty}  \frac{1}
{\sqrt{( i \bar{\omega  } _{n})^{2}-\bar{ \mu }^{2}  \bar{q}^{2}}} -
 \frac{1}{i \bar{\omega  } _{n}}
\label{eq6}
\end{eqnarray}
We have introduced:
\begin{eqnarray}
a _{0}(\bar{\mu },T) = \frac{1}{N _{0}V} - 2 \pi T  \: {\rm Re}  
\sum_{n=0} ^{  \omega _{c} } \frac{1}{\bar{\omega  }_{n}}
\label{eq7}
\end{eqnarray}
which is zero on the spinodal transition line (the line in the $ \bar{ \mu 
},T $ plane where the normal state becomes absolutely unstable against 
a transition toward a space independent order parameter) and is positive 
above it. At the FFLO transition we are looking at, we have $\Omega 
_{2}(q, \bar{ \mu },T)=0$. The actual transition corresponds to the 
largest possible $ \bar{ \mu }$ at fixed $T$. From Eq.(7) this 
corresponds to have the largest $ a _{0}(\bar{\mu },T) $. Hence from 
Eq.(5) we want to minimize $I(q, \bar{ \mu },T)$ with respect to $q$. 
At low temperature it is more convenient to express $I$ as:
\begin{eqnarray}
I(q, \bar{ \mu },T) = - \frac{1}{2}  \: {\rm Re} 
\int_{- \infty} ^{ \infty} \! d \omega  \: \tanh( \frac{\omega }{2T}) \, [ 
\frac{1}
{\sqrt{( \omega + \bar{\mu} )^{2}-\bar{ \mu }^{2}  \bar{q}^{2}}} -
 \frac{1}{ \omega + \bar{\mu} }]
\label{eq8}
\end{eqnarray}
where the integration contour runs actually infinitesimally above the real 
$ \omega $ axis. 

At  $T=0$ the integration is easily performed to give:
\begin{eqnarray}
I(q, \bar{ \mu },T) = {\rm Re} \ln ( 1+ \sqrt{1-\bar{q}^{2}}) - \ln 2
\label{eq9}
\end{eqnarray}
The minimum is reached for $ \bar{q} = 1 $ , in agreement with 
Shimahara \cite{shima1} and Burkhardt and Rainer \cite{br}, and $ I = 
- \ln 2 $ at this minimum. In this case, from Eq.(7), $ a _{0}(\bar{\mu 
},0) =  \ln(2  \bar{ \mu }/\Delta _{0})$ where $ \Delta _{0} = 2 \omega 
_{c} \exp(-1/ N _{0}V) $ is the zero temperature BCS phase gap (this 
corresponds to the value $ \bar{\mu }= \Delta _{0}/ 2 $ for the spinodal 
transition). This leads to $ \bar{\mu }= \Delta _{0} $ for the location of 
the FFLO transition, again in agreement with previous work. It is worth 
to note that, as already mentionned in the introduction, the location $ 
\bar{q} = 1 $ of the minimum corresponds to a singular point for $ I $ 
since we have explicitely $ I = \ln(\bar{q}/2)$ for $ \bar{q} > 1 $. 
While $ I $ itself is continuous, its derivative is discontinuous for $ 
\bar{q} = 1 $.

For $T \neq 0$ there is no singular behaviour and we find the value of $ 
\bar{q} $ giving the minimum   $I$ by writing that its derivative with 
respect to $ \bar{q} $ is zero. Integrating the result by parts leads to the 
condition:
\begin{eqnarray}
{\rm Re} \int_{- \infty} ^{ \infty} \! dy  \:  \frac{1}{\cosh ^{2} y} \,  
\frac{1+2ty}
{\sqrt{(1+2ty )^{2}-\bar{q}^{2}}} = 2
\label{leading}
\end{eqnarray}
where we have taken the new variable $ y = \omega / 2T $ and defined 
the reduced temperature $ t = T /  \bar{\mu}$. Only the ranges $ y > 
(\bar{q}-1)/2t \equiv a$ and $ y < - (\bar{q}+ 1)/2t $ contribute to the 
real part of the integral. Since at low $T$ we have $ \bar{q} \approx 1 
$, this last range will only give an exponentially small contribution 
because of the factor $ \cosh ^{-2} y $. Since this same factor makes $ | 
y |$ to be at most of order 3 , we can make at low $T$ to leading order 
$1+2ty \approx 1$ and $ (1+2ty )^{2}-\bar{q}^{2} \approx 2 ( 1 - 
\bar{q} + 2ty )$. It is then seen that we must have $  (\bar{q}-1)/2t \gg 
1$, because $  (\bar{q}-1)/2t \approx 1$ makes the left hand side of 
Eq.(10) much larger than unity at low $T$. This implies $ y \gg 1 $ and 
$ 2 \cosh y \approx \exp y$. The integral is then easily evaluated and 
Eq.(10) gives finally to leading order:
\begin{eqnarray}
\bar{q}-1 = \frac{t}{2} \,  \ln  \frac{\pi }{2t} 
\label{leading2}
\end{eqnarray}
Note that this result is in disagreemeent with the analysis given 
\cite{bul} by L. N. Bulaevskii. The reason for this discrepancy is 
discussed in details in Appendix A. In particular we rederive in this 
appendix our Eq. (\ref{leading}) from the starting equation of Ref. 
\cite{bul}.  

\begin{figure}[h]
\begin{center}
\includegraphics[width=0.6\textwidth,height=0.3
\textheight]{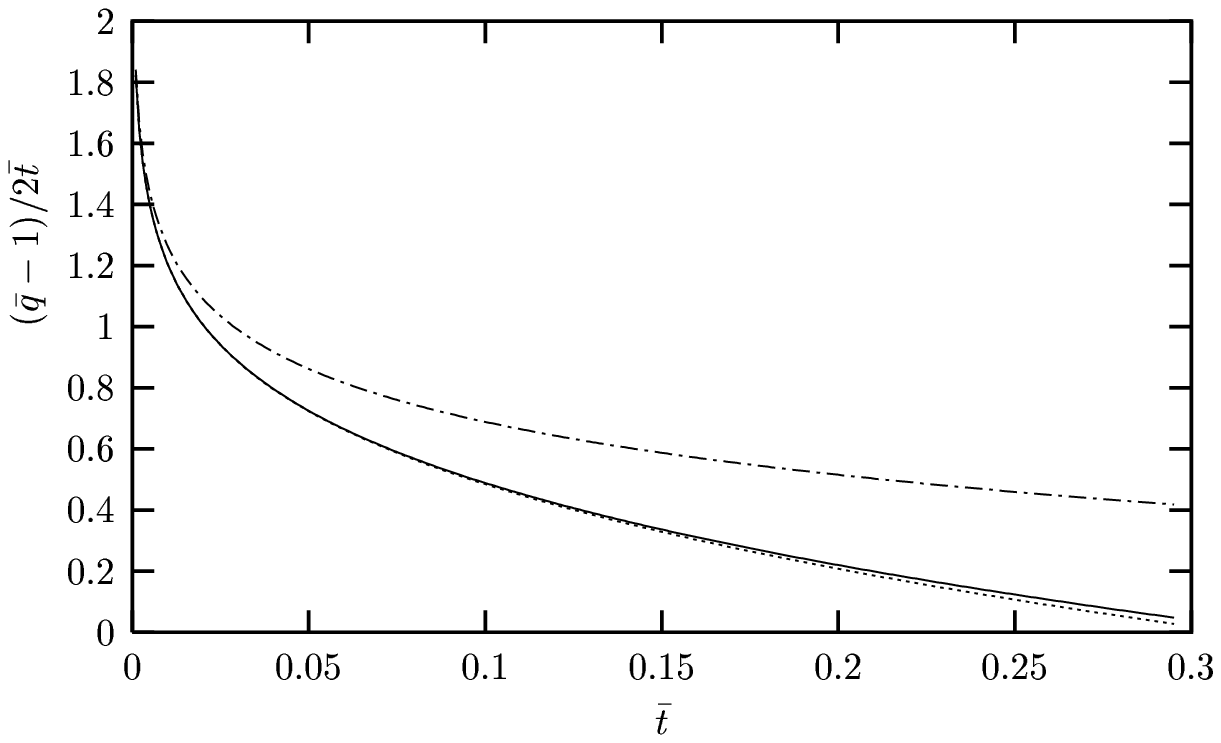}
\includegraphics[width=0.6\textwidth,height=0.3
\textheight]{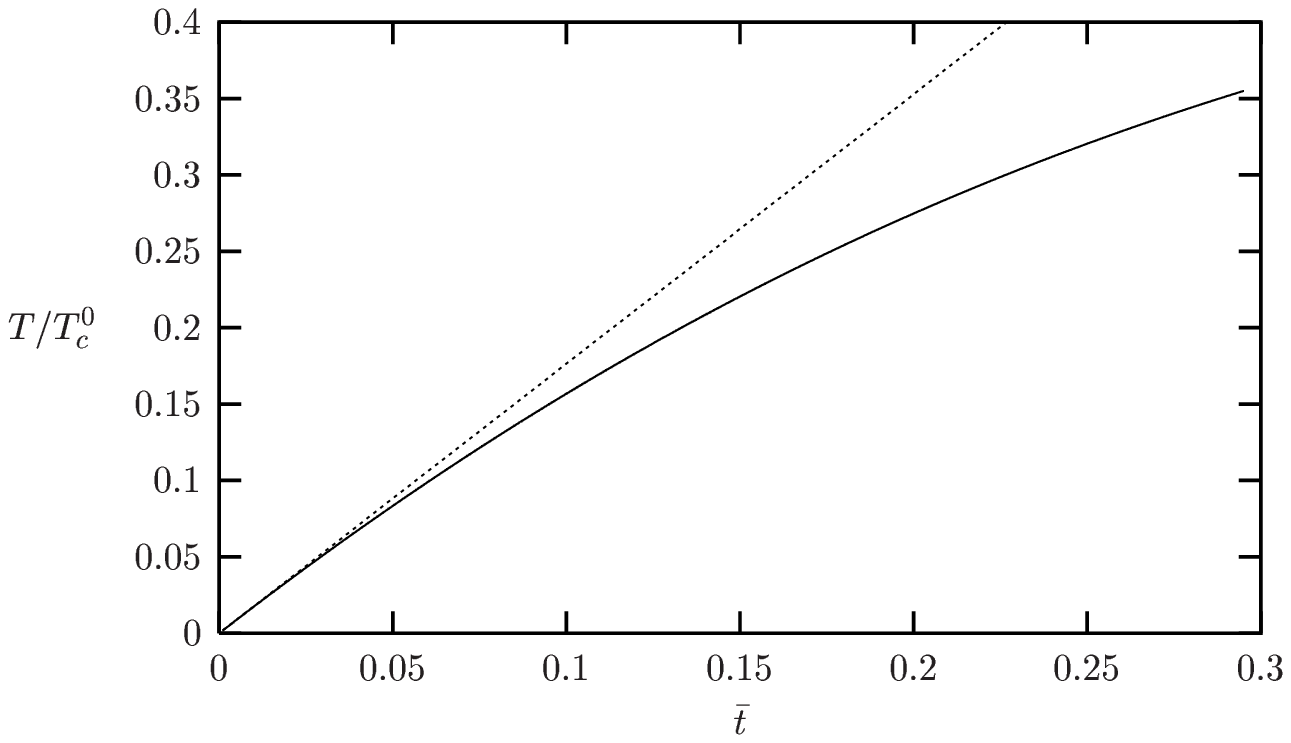}
\caption{$a=(\bar{q}-1)/2t$ from the leading low temperature 
approximation Eq.(\ref{leading2}) (dashed-dotted line), the low 
temperature expansion Eq. (\ref{alowtemp}) (dotted line) and the full 
numerical calculation (full line), as a function of $\bar{t} = t/\bar{q}$. 
The lower panel gives the relation between $\bar{t}$ and $ T/ T_{c0}$ (full
line), the dotted line is the low temperature limiting behaviour 
$T / T_{c0} = \bar{t} \Delta _{0}/T_{c0} = 1.76 \bar{t}$.}
\label{fig0}
\end{center}
\end{figure}

A more complete low temperature expansion
can be fairly easily extracted from Eq.(\ref{leading}). As previously 
seen,
the range $ y >  a $ in the integration
is sufficient since the other integration range gives an exponentially
small term. With the change of variable
$y = u^2 + a $, Eq.(\ref{leading}) leads to:
\begin{equation}\label{leading3}
\int_0^{+\infty} du \frac{1+ 2 \bar{t} u^2}{\cosh^2(u^2+a) 
\sqrt{1+\bar{t}
u^2}}  = 2 \sqrt{\bar{t}}
\end{equation}
where we have defined $\bar{t} = t/\bar{q}$.
Neglecting terms of order $\bar{t} ^{2}$ Eq. (\ref{leading3}) can be 
written as:
\begin{equation}\label{devel}
\int_0^{+\infty} \frac{ du}{\cosh^2(u^2+a)} + \frac 3 2
\bar{t}   \int_0^{+\infty} \frac{u^2 du}{\cosh^2(u^2+a)} = 2 
\sqrt{\bar{t}}
\end{equation}
From Eq.(\ref{leading2}) the leading order for $a$ is $a_0 = (1/4) \ln 
(\pi / 2 \bar{t}) \gg 1 $.
In the second integral in Eq.(\ref{devel}) we can replace $a$ by $a_0$ 
since $\exp(-2 a_0) = \sqrt{2 \bar{t}/\pi} \ll 1$. Setting $ a = a_0 + 
\delta a$, we obtain an expansion of $ \delta  a$ in powers of $\bar{t} 
^{1/2}$ by expanding the first integral up to second order in powers of 
$ \delta a$:
\begin{equation}
- \frac{4}{\sqrt{\pi} } \bar{t} + \frac{4 \sqrt{3}}{\pi} \bar{t}^{3/2} + 
\delta a \left(- 4 \bar{t} ^{1/2} + \frac{16}{\sqrt{\pi}}
\bar{t} \right) + 4 (\delta a)^2 \bar{t} ^{1/2} + \frac{3}{4} 
\bar{t}^{3/2} = 0
\end{equation}
valid up to the order $ \bar{t} ^{3/2}$. Solving this equation order by 
order, we find the following expansion for $a$:
\begin{equation}
a = \frac{1}{4} \ln\left(\frac{ \pi}{2 \bar{t}}\right)
- \sqrt{ \frac{\bar{t}}{\pi}  }
+ \bar{t} \left( \frac{3}{16} +
\frac{\sqrt{3}}{\pi} - \frac{3}{\pi} \right) + {\cal O}
( \bar{t}^{3/2}  ).
\label{alowtemp}
\end{equation}
This expression leads to a marked improvement when it is compared to 
the straight 
numerical evaluation of the Matsubara sums in Eq.(6). This is seen in 
Fig.\ref{fig0} where we have plotted the optimal $a=(\bar{q}-1)/2t$ 
from straight numerical calculation, as well as its leading low 
temperature approximation and the one resulting from the expansion 
Eq.(\ref{alowtemp}). We give also for convenience, in the lower panel, the 
dependence of our reduced temperature $ \bar{t}$ 
as a function of the ratio $ T /  
T_{c0}$, where $T_{c0}$ is the standard BCS critical temperature, found for
$\bar{\mu  }=0$. In the low temperature limit $ T \rightarrow 0 $, 
we have merely $ \bar{q}=1$ and $ \bar{\mu }= \Delta _{0} $, hence
$ \bar{t}= T/\Delta _{0} $ and the dotted straight line in the lower 
panel gives this limiting behaviour $ T /
T_{c0} = \bar{t} \Delta _{0}/T_{c0}$.

\section{FOURTH ORDER TERM}
\subsection{Leading behaviour}

The second order term in the free energy gives the same transition line 
for all the combinations of plane waves. The selection among the 
various possible order parameters will be made by the fourth order term 
in the free energy expansion, as pointed out by Larkin and 
Ovchinnikov. The selected state will correspond to the lowest fourth 
order term. So we turn now to this fourth order term. Its general 
expression is given by Eq.(2) with \cite{larkov} :
\begin{eqnarray}
N _{0} \Omega _{4}({{\bf  q}}_{1},{{\bf  q}}_{2},{{\bf  
q}}_{3},{{\bf  q}}_{4},
\bar{ \mu },T) =  T  \sum_{n,k}  \bar{G}({\bf k}) G({\bf k}+{\bf 
q}_{1})
\bar{G}({\bf k}+{\bf q}_{1}-{\bf q}_{4}) G({\bf k}+{\bf q}_{2})
\label{eq12}
\end{eqnarray}
where we have used already momentum conservation ${\bf  q}_{1} + 
{\bf  q}_{3} = {\bf  q}_{2} + {\bf  q}_{4}$. Since all the 
wavevectors ${\bf  q}_{i}$ have the same modulus given by Eq.(11), 
this momentum conservation implies in our 2D situation either  
\cite{larkov} ${\bf  q}_{1} = {\bf  q}_{2}$ together with $ {\bf  
q}_{3} = {\bf  q}_{4}$, or equivalently ${\bf  q}_{1} = {\bf  
q}_{4}$ together with $ {\bf  q}_{3} = {\bf  q}_{2}$. Moreover one 
has the additional possibility ${\bf  q}_{1} = - {\bf  q}_{3}$ together 
with $ {\bf  q}_{2} = - {\bf  q}_{4}$.  The two first possibilities lead 
to a same coefficient called $ J(\alpha _{{\bf  q}_{1},{\bf  q}_{3}})$ 
by LO where $\alpha _{{\bf  q}_{1},{\bf  q}_{3}}$ is the angle 
between ${\bf  q}_{1}$ and ${\bf  q}_{3}$. Similarly the last 
possibility leads to a coefficient $ \tilde{J}(\alpha _{{\bf  q}_{1},{\bf  
q}_{2}})$. Explicitely the fourth order term in the free energy (last 
term of the r.h.s. of Eq.(2)) becomes  \cite{larkov} : 
\begin{eqnarray}
\frac{1}{2} \sum_{i,j} (2- \delta _{{\bf  q}_{i},{\bf  q}_{j}}) |  \Delta 
_{{\bf  q}_{i}}| ^{2}  |  \Delta _{{\bf  q}_{j}}| ^{2} J(\alpha _{{\bf  
q}_{i},{\bf  q}_{j}}) + (1-\delta _{{\bf  q}_{i},{\bf  q}_{j}}-\delta 
_{{\bf  q}_{i},-{\bf  q}_{j}})  \Delta  _{{\bf  q}_{i}}  \Delta _{-{\bf  
q}_{i}}  \Delta ^{*} _{{\bf  q}_{j}}  \Delta  ^{*}_{-{\bf  q}_{j}}  
\tilde{J} (\alpha _{{\bf  q}_{i},{\bf  q}_{j}})
\label{fourthorder}
\end{eqnarray}
with, after a change of variable in the summation:
\begin{eqnarray}
N _{0} J(\alpha _{{\bf  q}_{1},{\bf  q}_{2}})  =  T  \sum_{n,k}  
\bar{G}({\bf k}+{\bf  q}_{1}) \bar{G}({\bf k}+{\bf  q}_{2}) G 
^{2}({\bf k})
\label{eq14}
\end{eqnarray}
and
\begin{eqnarray}
N _{0} \tilde{J} (\alpha _{{\bf  q}_{1},{\bf  q}_{2}})  =  T  
\sum_{n,k}  \bar{G}({\bf k}) \bar{G}({\bf k}+{\bf  q}_{1}+{\bf  
q}_{2}) G({\bf k}+{\bf  q}_{1}) G({\bf k}+{\bf  q}_{2})
\label{eq15}
\end{eqnarray}
When the $ \xi _{{\bf  k}}$ integration is performed, taking into 
account $  \xi _{{\bf  k}+{\bf  q}} \approx   \xi _{{\bf  k}}+ {\bf  
k}.{\bf  q}/m $ since $ q \ll k _{F} $, one finds:
\begin{eqnarray}
J(\alpha _{{\bf  q}_{1},{\bf  q}_{2}}) = 4 \pi T   \: {\rm Im} 
\sum_{n=0} ^{ \infty}
<  \frac{1}{(2i\bar{\omega  } _{n}+{\bf  k}.{\bf  q} _{1}/m) 
^{2}(2i\bar{\omega  } _{n}+{\bf  k}.{\bf  q} _{2}/m) } + 1 
\leftrightarrow 2 >
\label{eq16}
\end{eqnarray}
where the bracket means the angular average, and similarly:
\begin{eqnarray}
\tilde{J} (\alpha _{{\bf  q}_{1},{\bf  q}_{2}}) = -16 \pi T   \: {\rm 
Im} \sum_{n=0} ^{ \infty}
<  \frac{ i\bar{\omega  } _{n}}{[(2i\bar{\omega  } _{n})^{2}-({\bf  
k}.{\bf  q} _{1}/m) ^{2}].[(2i\bar{\omega  } _{n})^{2}-({\bf  
k}.{\bf  q} _{2}/m) ^{2}]} >
\label{eq17}
\end{eqnarray}
on which it is clear that $\tilde{J} (\pi - \alpha ) = \tilde{J} (\alpha )$.

Let us first consider $ J(\alpha)$. As in the preceding section when 
going from Eq.(6) to Eq.(8), it is more convenient to transform the sum 
over Matsubara frequencies into an integral on the real frequency axis. 
When one furthermore performs a by parts integration over the 
frequency $ \omega $, one gets from Eq.(\ref{eq16}):
\begin{eqnarray}
4 \bar{ \mu }^{2} J(\alpha) = - {\rm Re} 
\int_{- \infty} ^{ \infty} \! d \omega  \: \frac{1}{\cosh ^{2}( \omega 
/2T)} 
\int_{0} ^{2 \pi }  \! \frac{d \theta }{2 \pi } \frac{1}{[\bar{q} \cos ( 
\theta - \alpha /2)-(1+ \omega / \bar{ \mu })].[\bar{q} \cos ( \theta + 
\alpha /2)-(1+ \omega / \bar{ \mu })]}
\label{eq18}
\end{eqnarray}
where as above $ \bar{q} = q k_{F} / 2m \bar{ \mu } $ and the 
integration contour runs again infinitesimally above the real $ \omega $ 
axis. Here in the angular integration we have taken the reference axis 
bisecting the angle between ${\bf  q}_{1}$ and ${\bf  q}_{2}$. This 
angular integration is performed by residues, taking $ \exp(i \theta ) $ as 
a variable. For $Y = (1+ \omega / \bar{ \mu })/\bar{q}$ this leads to:
\begin{eqnarray}
\int_{0} ^{2 \pi }  \! \frac{d \theta }{2 \pi } \frac{1}{[\cos ( \theta - 
\alpha /2)-Y].[\cos ( \theta + \alpha /2)-Y]} = \frac{Y}{(Y ^{2}- 
\cos^{2} ( \alpha /2)) \sqrt{ Y ^{2}- 1}} 
\label{eq19}
\end{eqnarray}
where the cut in the determination of the square root has to be taken on 
the positive real axis, as it is clear when one considers the case of very 
large $| Y |$. When this result is inserted in Eq.(\ref{eq18}) and $Y$ is 
taken as new variable, one finds, introducing again the reduced 
temperature $ t = T /  \bar{\mu}$:
\begin{eqnarray}
16 t \bar{ \mu }^{2} \bar{q} J(\alpha) = - {\rm Re} 
\int_{- \infty} ^{ \infty} \! dY  \: \frac{1}{\cosh ^{2}(( \bar{q}Y-1) 
/2t)} 
\frac{Y}{(Y ^{2}- Y _{1}^{2}) \sqrt{ Y ^{2}- 1}}
\label{eq20}
\end{eqnarray}
with $ Y _{1} =  \cos ( \alpha /2) $.

Up to now we have made no approximation in our calculation and the 
result is valid at any temperature. Let us now focus on the low 
temperature regime $ t \ll 1$. Because of the factor $\cosh ^{-2}(( 
\bar{q}Y -1) /2t)$, only the vicinity of $Y=1$ will contribute. 
Moreover only the half circle contour around the pole $Y= Y _{1}$ and 
the range $ Y > 1 $ contribute to the real part. In this last domain we 
have $ (\bar{q}Y -1) /2t > (\bar{q}-1) /2t  \gg 1 $ from our result for 
$\bar{q}$ Eq.(11), so we can again simplify the hyperbolic cosine into 
an exponential (although this is not in practice a good approximation 
numerically). With the further change of variable $Y= 1+t v ^{2}/ 
\bar{q}$ in the resulting integral, we find to leading order:
\begin{eqnarray}
16 t \bar{ \mu }^{2} J(\alpha) = \frac{\pi }{\alpha } \frac{1}{\cosh 
^{2}[(1/4) \ln  \frac{\pi }{2t} - \beta ^{2} /2]} - \frac{8}{\sqrt{ 
\pi}(1+ \cos ( \alpha /2)) } 
\int_{0} ^{ \infty} \! dv  \: \frac{\exp(-v ^{2})}{ v ^{2}+\beta ^{2}} 
\label{Jalpha}
\end{eqnarray}
where we have substituted explicitely the result Eq.(11) for $\bar{q}$ 
and have set $ \beta ^{2} = (1-\cos ( \alpha /2)) /t $. The integral in this 
result can not be further simplified in general and is related to parabolic 
cylinder functions. Let us consider now some important limiting cases 
for this result. First we take at fixed $ \alpha $ the limit $ T \rightarrow 
0$. This implies $ \beta ^{2} \rightarrow \infty$, the first term goes to 
zero and the integral is easily calculated in this limit, leading to:
\begin{eqnarray}
J(\alpha) = -  \frac{1 }{4   \mu ^{2}} \frac{1}{\sin^{2} ( \alpha /2)} 
\label{Jlargealpha}
\end{eqnarray}
On the other hand if at fixed $ T $ we take the limit $ \alpha \rightarrow 
0$, we have $ \beta ^{2} \rightarrow 0$. The limiting behaviour of the 
integral is $ \pi /(2 \beta ) - \pi ^{1/2}$ as can be obtained through a by 
parts integration, the dominant divergent contribution from the two 
terms cancels out and we are left with \cite{remark}:
\begin{eqnarray}
J(\alpha) =  \frac{1 }{4   \mu ^{2}} \frac{1}{t} 
\label{eq23}
\end{eqnarray}
which goes naturally to infinity for $ T \rightarrow 0$. These two limits 
can be obtained more rapidly by making the proper simplifications from 
the start of the calculation Eq.(\ref{eq18}).

The two limiting cases which we have just considered show that, at low 
temperature, $ J(\alpha)$ has a quite remarkable behaviour. For most of 
the range it is negative as it can be seen from Eq.(\ref{Jlargealpha}) and 
it even goes to large negative values when $ \alpha $ gets very small. 
This tends to favor states with small angle between wavevectors, as we 
will see below. On the other hand for $ \alpha =0$ or very small $ 
J(\alpha)$ is positive and very large, as it results from Eq.(\ref{eq23}). 
Clearly at low $T$ the interesting range is the small $ \alpha $ domain 
where we can write $ \beta =  \alpha /(8t) ^{1/2} $. Surprisingly $ 
J(\alpha)$ first starts to increase strongly from its $ \alpha = 0$ value 
before going down to very negative values. This can be seen simply by 
looking at the specific point $\beta ^{2} = (1/2) \ln (\pi /2t) $ for which 
the second term is negligible and which gives to dominant order $ 
J(\alpha)= \pi /(32  \mu ^{2}) t ^{-3/2}(\ln (\pi /2t)) ^{-1/2}$, even 
more diverging for $ T \rightarrow 0$ than the $ T=0$ value. The 
integral of $ J(\alpha)$ over $ \alpha $ can also be analytically 
evaluated, and it shows that the strong positive peak at small $ \alpha $ 
dominates over the negative contribution from the rest of the range. 
This quasi-singular behaviour of $ J(\alpha)$ is summarized in 
Fig.\ref{fig1} where we have plotted $ J(\alpha)$ for various 
temperatures.

We perform now the same kind of treatment for $\tilde{J} (\alpha )$. 
One goes again from Eq.(\ref{eq17}) to an integration on the real 
frequency axis. However there is no integration by parts, and the 
angular average to be calculated is somewhat more complicated. It can 
nevertheless be performed by the same method and gives explicitely, 
with the same variable $Y = (1+ \omega / \bar{ \mu })/\bar{q}$ as 
above:
\begin{eqnarray}
\int_{0} ^{2 \pi }  \! \frac{d \theta }{2 \pi } \frac{1}{[\cos ^{2} ( \theta 
- \alpha /2)-Y^{2}].[\cos ( \theta + \alpha /2) ^{2}-Y^{2}]} = 
\frac{1}{2Y  \sqrt{ Y ^{2}- 1}} ( \frac{1}{Y ^{2}- \cos^{2} ( \alpha 
/2)}+ \frac{1}{Y ^{2}- \sin^{2} ( \alpha /2) }) 
\label{eq24}
\end{eqnarray}
The second term is obtained from the first one by the change $ \alpha 
\rightarrow \pi  - \alpha $. This leads to $\tilde{J} (\alpha) = 
\tilde{J}_o(\alpha) + \tilde{J}_o(\pi- \alpha)$ with:
\begin{eqnarray}
8 \bar{ \mu }^{2} \bar{q} ^{2}\tilde{J}_o (\alpha) = - {\rm Re} 
\int_{- \infty} ^{ \infty} \! dY  \:\tanh ((\bar{q}Y-1) /2t) \frac{1}{ 
\sqrt{ Y ^{2}- 1}} 
 \frac{1}{ Y ^{2}- \cos^{2} ( \alpha /2)}
\label{eq25}
\end{eqnarray}
Here we have contributions coming from half circles around the poles at 
$ Y = \pm  \cos ( \alpha /2) = \pm Y _{1}$ and contributions from the 
two domains $ Y > 1$ and $ Y < -1$. Neglecting terms which are 
exponentially small  in the low temperature limit, we can replace the 
hyperbolic tangent by $-1$ in the $Y<-1$ domain and for the $Y = - Y 
_{1}$ pole. Gathering similar contributions this gives:
\begin{eqnarray}
8 \bar{ \mu }^{2} \bar{q} ^{2}\tilde{J}_o (\alpha) = \frac{ \pi }{\sin 
\alpha }
\tanh ((\bar{q}Y _{1}-1) /2t) +  \int_{1} ^{ \infty} \! dY  \:
 \frac{1 - \tanh ((\bar{q}Y-1) /2t)}{ (Y ^{2}- \cos^{2} ( \alpha /2)) 
\sqrt{ Y ^{2}- 1}}
+ \frac{2 \alpha - \pi }{\sin \alpha } 
\label{eq26}
\end{eqnarray}
where we have used the fact that, without the $ \tanh ((\bar{q}Y-1) 
/2t)$ term, the integral can be performed exactly to give $(\pi - \alpha )/ 
\sin \alpha $. We have taken advantage of this to have the factor $ 1 -  
\tanh ((\bar{q}Y-1) /2t)$ which is going rapidly to zero for large $Y$. 
This will be of use when we consider below temperature corrections. 
The last term in Eq.(\ref{eq26}) disappears in the combination 
$\tilde{J}_o(\alpha) + \tilde{J}_o(\pi- \alpha)$, so we omit it from now 
on. 

Looking now for the dominant contribution at low temperature, we 
simplify the hyperbolic tangent in the range $ Y > 1$ since its argument 
$ (\bar{q}Y-1) /2t $ is large and positive. So one finds expressions 
which are similar to the one encountered in the calculation of $ 
J(\alpha)$. Moreover, because of the relation $\tilde{J} (\pi - \alpha ) = 
\tilde{J} (\alpha )$, we can restrict ourselves to the case $ \alpha \leq \pi 
/2$. In this case the expression obtained by the replacement $ \alpha 
\rightarrow \pi  - \alpha $ is very simple since the corresponding value 
of $ \beta ^{2} = (1-\cos ( \alpha /2)) /t $ is always very large at low T. 
This leads finally to: \begin{eqnarray}
8 \bar{ \mu }^{2} \tilde{J}(\alpha) = - \frac{2 \pi }{\sin \alpha } ( 1 - 
\frac{1}{1+ (2t / \pi ) ^{1/2} \exp( \beta ^{2})}) + \frac{2}{\sqrt{ 
\pi}} 
\int_{0} ^{ \infty} \! dv  \: \frac{\exp(-v ^{2})}{ v ^{2}+\beta ^{2}} 
\label{eq27}
\end{eqnarray}
where we have taken into account that the second term is only 
significant when $ \alpha $ is small, so we have made $ \alpha = 0$ in 
its prefactor.

In the limit $ T \rightarrow 0$ with fixed $ \alpha $, we have $ \beta 
^{2} \rightarrow \infty$ and one gets merely:
\begin{eqnarray}
\tilde{J} (\alpha) = -  \frac{ \pi  }{4   \mu ^{2}} \frac{1}{\sin \alpha } 
\label{jtildelim}
\end{eqnarray}
Again if we take $ \alpha \rightarrow 0$ at fixed $ T $, $ \beta ^{2} 
\rightarrow 0$, the dominant divergent contribution cancels and we 
find:
\begin{eqnarray}
\tilde{J} (\alpha) =  - \frac{1 }{4   \mu ^{2}}
\label{eq29}
\end{eqnarray}
These two limits can be obtained again more directly. From these cases 
we can guess that $\tilde{J} (\alpha)$ is always negative. This is seen 
on Fig. \ref{fig2} where $ \tilde{J} (\alpha)$ has been plotted. 
However, in the same way as $J( \alpha )$, it has also a singular 
behaviour at small $ \alpha $. While for $ \beta \approx  \alpha /(8t) 
^{1/2} \gg 1$ it diverges as $ - \pi /(4   \mu ^{2} \alpha )$, it goes to 
the finite value $  - 1/(4   \mu ^{2})$ for $ \alpha =0$. We note that the 
divergent behaviour in $ \alpha ^{-1}$ is weaker than the one in $ 
\alpha ^{-2}$ found for $J( \alpha )$ . Similarly $J(0) \gg |\tilde{J}(0) 
|$ at low $T$. So $J( \alpha )$ will play the dominant role and 
$\tilde{J} (\alpha)$ will only give a subdominant contribution.

\subsection{Temperature corrections}

The expressions Eq.(\ref{eq20}) and Eq.(\ref{eq27}) found above 
respectively for $J(\alpha)$
and $\tilde{J}(\alpha)$ are the leading terms at low temperature. While 
containing the dominant physical behaviour, we do not expect them to 
be so good quantitatively at intermediate temperature. It is actually 
possible to improve these analytical results markedly in this respect at 
the price of a slight complication by including the first two terms in an 
expansion in powers of $ \bar{t}^{1/2}$. This is the similar to what 
we have done at the end of section II for the second order term, and we 
follow here the same procedure, together  with the steps we have just 
taken above for the calculation of $J(\alpha)$ and $\tilde{J}(\alpha)$.

\begin{figure}[h]
\begin{center}
\includegraphics[width=0.6\textwidth,height=0.3
\textheight]{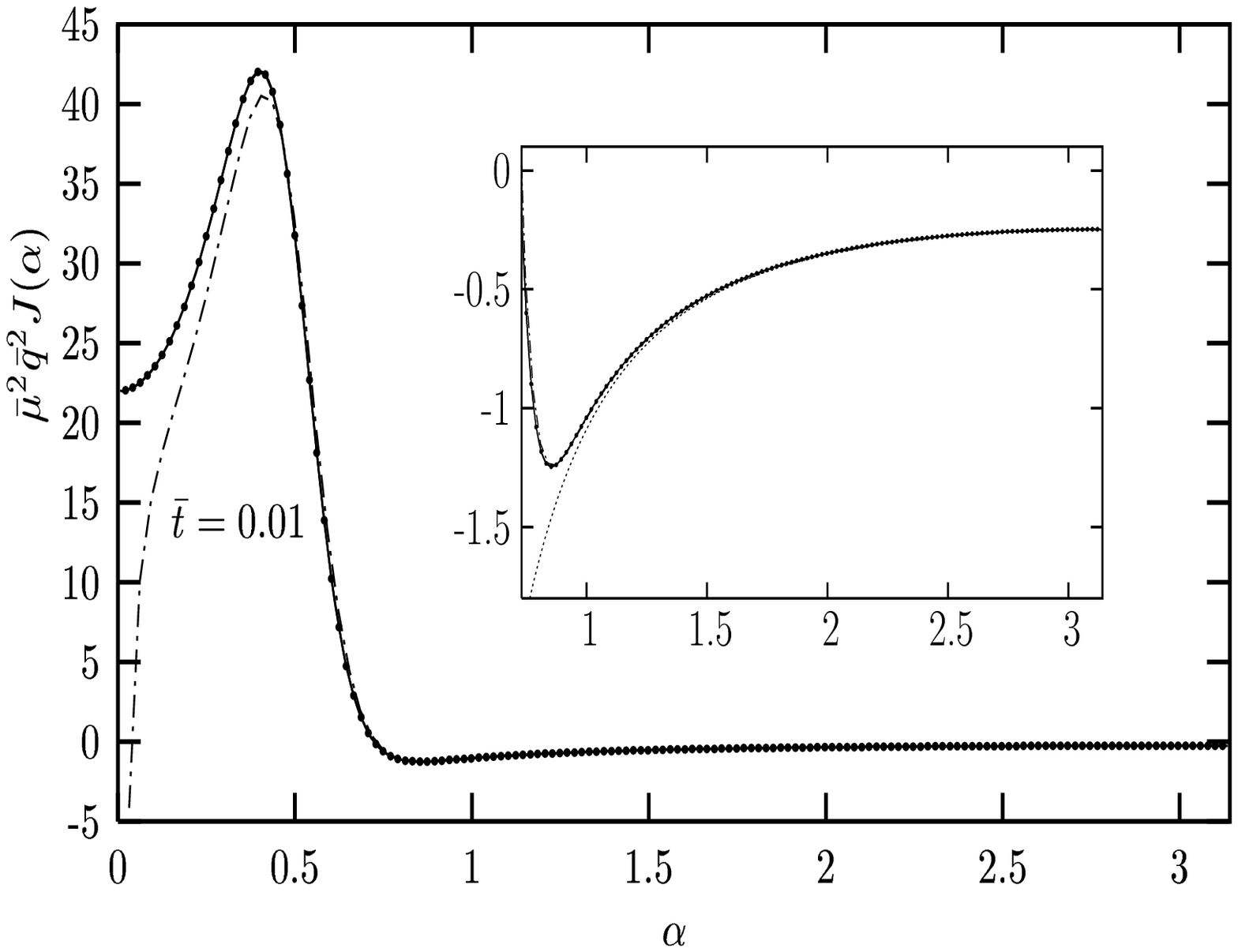}
\includegraphics[width=0.6\textwidth,height=0.3
\textheight]{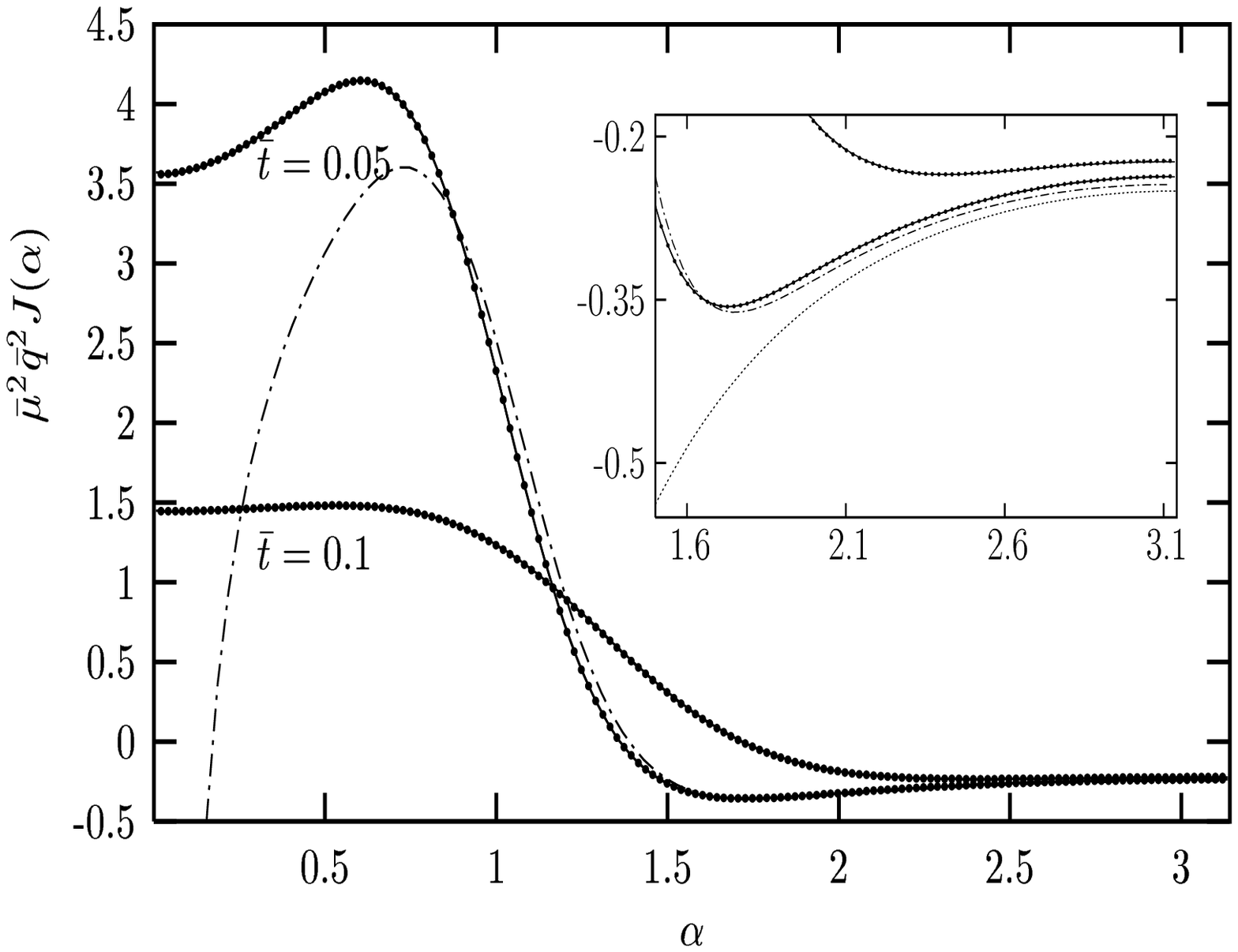}
\caption{$J(\alpha)$ for various values of
the reduced temperature $\bar{t} = T/(\bar{q} \bar{\mu}) = 0.01$,
$0.05$ and $0.1$. The lines are calculated using
our low temperature expressions for $J(\alpha)$
and $a$, Eqs.(\ref{eqJ}) and (\ref{alowtemp}), while
the points correspond to the exact result numerical summation over 
Matsubara frequencies. The dashed line is the  asymptotic behaviour for 
$J(\alpha)$ in the $T \to 0$
limit Eq.(\ref{Jlargealpha}). The dashed-dotted lines are the leading
order low temperature result Eq.(\ref{Jalpha})}
\label{fig1}
\end{center}
\end{figure}

We first consider first $J(\alpha)$ and start from the exact 
Eq.(\ref{eq20}). Since the contribution from the domain $ Y < -1$ is 
again exponentially small in
the low temperature regime, we keep only the half circle contour around 
$Y_1$
and the $Y> 1$ domain. The integral appearing in Eq.(\ref{eq20})
can be split in two terms by:
\begin{equation}
\frac{Y}{Y^2-Y_1^2} = \frac 1 2 \left( \frac{1}{Y-Y_1} + 
\frac{1}{Y+Y_1}
\right).
\end{equation}
Therefore $J(\alpha)$ can be written as $J(\alpha) = J_o(\alpha) + J_o(2 
\pi-
\alpha)$ and we concentrate on the calculation of  $J_o$. With the same 
notations $a=(\bar{q}-1)/2t $, $\bar{t} = t/\bar{q}$ and the change of 
variable $ (\bar{q} Y - 1)/2 t = a + u^2 $,
Eq.(\ref{eq20}) leads to:
\begin{equation}\label{Jo}
32 \bar{t} \bar{\mu}^2 \bar{q}^2 J_o(\alpha) = - \bar{t}^{-1/2}
\int_0^{\infty} \frac{du}{\cosh^2(u^2+a) }
 \frac{1}{ (u^2+\bar{\beta}^2/2) \sqrt{1+\bar{t} u^2}}
+  \frac{\pi}{\sin(\alpha/2) \cosh^2(a-
\bar{\beta}^2/2)}
\end{equation}
where we have set $\bar{\beta}^2 = (1-\cos(\alpha/2))/ \bar{t}$. The 
first term in the r.h.s. of Eq.(\ref{Jo}) is expanded for low $t$ by 
$(1+\bar{t} u^2)^{-1/2} \simeq
1 - \bar{t} u^2 /2$ and the resulting temperature correction is calculated
to lowest order, replacing $a$ by $a_0 = \ln(\pi/2\bar{t})/4$. This gives
our following final expression, which has to be used together with 
Eq.(16) for $a$ and $\bar{q}$:
\begin{equation}
32 \bar{t} \bar{\mu}^2 \bar{q}^2 J_o(\alpha) = - \bar{t}^{-1/2}
\int_0^{\infty} \frac{du}{\cosh^2(u^2+a) } \frac{1} 
{u^2+\bar{\beta}^2/2 }
+  \frac{\pi}{\sin(\alpha/2) \cosh^2(a-
\bar{\beta}^2/2)} + \frac{2 \bar{t}}{\sqrt{\pi}} \int_0^{\infty} dv 
\frac{v^2
\exp(-v^2)}{v^2+\bar{\beta}^2}
\label{eqJ}
\end{equation}
The second term in the r.h.s of this equation is only relevant for angles
that are close to zero. Otherwise, it gives an exponentially small 
contribution
in the low temperature limit. In the expression obtained by the change 
$\alpha
\to 2 \pi -\alpha$, this term is exponentially small
in any case and can be forgotten. The first and the third term can be
calculated more explicitly for angles $ \alpha $ which are not close to 
zero (this is in particular the case for $J_o (2 \pi - \alpha)$), which 
implies $\bar{\beta} \to + \infty$. For the first term in particular we can 
make use of our results in section II for the temperature expansion 
where similar terms were found. This leads finally in this regime to:
\begin{equation}
16 \bar{\mu}^2 \bar{q}^2 J_o(\alpha) = - \frac{2}{1-\cos(\alpha/2)}
+ \bar{t} \: \frac{2-\cos\alpha/2}{(1-\cos(\alpha/2))^2}
\end{equation}
For various temperatures, we compare in Fig. \ref{fig1} these low
temperature expressions with the exact calculation of $J(\alpha)$, from 
the direct numerical summation in Eq.(\ref{eq16}) over Matsubara 
frequencies (with $\bar{q}$ obtained by the numerical minimisation
of $I(q, \bar{ \mu },T)$ given by Eq.(\ref{eq6})).
We see that, at this level of accuracy, they agree remarkably well up to 
rather high temperatures.

We turn now to a similar calculation for $\tilde{J}(\alpha)$ and take up 
from Eq.(\ref{eq26}).
We have already pointed out that, since we can restrict ourselves to $0 
< \alpha
< \pi /2$, the replacement $\alpha \to \pi - \alpha$ gives a
value of $\bar{\beta}^2= (1-\cos(\alpha/2))/\bar{t}$ which is always
large at low temperature. Therefore, in this replacement, the integral in 
Eq.(\ref{eq26})
can be calculated to lowest order in $\bar{t}$
and gives $ 2 \bar{t} / \cos^2(\alpha/2)$. In the same way in the first 
term the hyperbolic tangent can be replaced by -1 in this substitution. 
With the same change of variable as for $J(\alpha)$, we have finally:
\begin{equation}
8 \bar{ \mu }^{2} \bar{q} ^{2}\tilde{J}_o (\alpha) = - \frac{ \pi }{\sin 
\alpha }
(1- \tanh(a -\bar{\beta}^2/2)) +  \frac{2 \bar{t}}{\cos^2(\alpha/2)}
+ \bar{t} ^{-1/2} \int_0^{\infty} du \frac{1-\tanh(u^2+a)}
{ ( 1+\cos(\alpha/2) + 2 \bar{t} u^2)(u^2+\bar{\beta}^2/2) 
\sqrt{1+\bar{t} u^2}}
\label{Jtilde}
\end{equation}
As for the calculation of $J(\alpha)$, we can have a more explicit result 
by expanding the denominator in the integral to first order in powers of 
$\bar{t}$ and computing the resulting temperature correction to lowest 
order by replacing $a$ by $a_0$. But we will not write the cumbersome 
resulting formula.

\begin{figure}[h]
\begin{center}
\includegraphics[width=0.6\textwidth,height=0.3
\textheight]{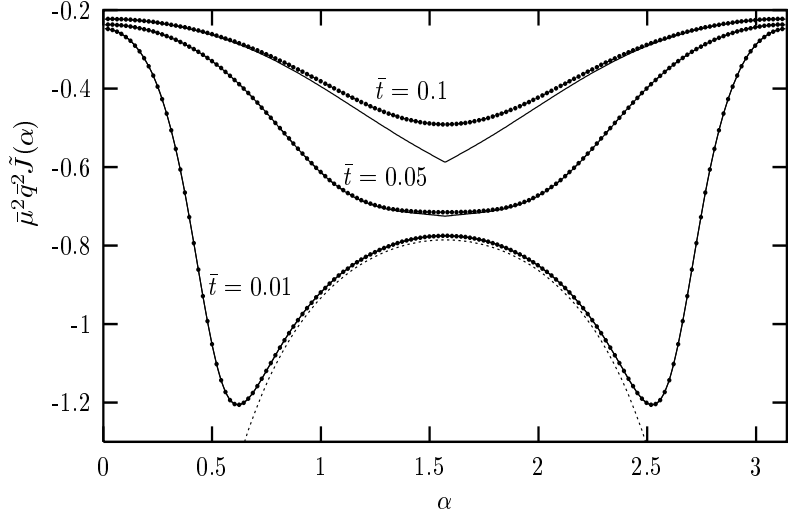}
\caption{$\tilde{J}(\alpha)$ plotted for various values of
the reduced temperature $\bar{t} = T/ (\bar{q} \bar{\mu}) = 0.01$,
$0.05$ and $0.1$. The lines correspond to the low
temperature expressions for $\tilde{J}(\alpha)$
and $a$, Eqs. (\ref{Jtilde}) and (\ref{alowtemp}).
while the dots show the exact results from numerical summation over 
Matsubara frequencies. The dashed line is the  asymptotic
behaviour Eq.(\ref{jtildelim}).}\label{fig2}
\end{center}
\end{figure}

Nevertheless, as done previously for $J(\alpha)$, we compare in Fig.
\ref{fig2} our low
temperature expressions with the exact calculation of $\tilde{J} 
(\alpha)$,
$\bar{q}$ and $\bar{\mu}$.
The discrepancy appears only above $\bar{t} \simeq 0.05$ in the vinicity
of $ \alpha  = \pi  /2$ and is 
mainly due
to the fact that it becomes inaccurate to consider that $ \beta ^{2}$ 
is large in calculating
$\tilde{J}_o(\pi- \alpha)$, as we have done (see above Eq.(\ref{eq27})).
However our low temperature expressions are already sufficient to describe 
the cascade.

\section{Cascade of order parameter structures}
\subsection{Ingredients responsible for the cascade}

The second order term in the expansion of the free energy $\Omega$ 
gives
the location of the transition $ \bar{ \mu  }(T)$ and the  optimum
wavevectors modulus $  \bar{q}$ entering the order
parameter $\Delta ({\bf r})$ but it can not distinguish
between different order parameter structures, because from
Eq.(\ref{freeener}) the  Fourier
components $\Delta_q$ of the order parameter
$\Delta({\bf r})$ are  decoupled in the second order term.
The resulting degeneracy is lifted by the fourth order term
when one goes slightly into the superfluid phase. This is the
basis of the analysis of Larkin and Ovchinnikov \cite{larkov} and
of Shimahara \cite{shima2}. So to speak, the wavevectors directions 
are independent at the level of the second order term while the fourth 
order one provides an effective interaction between these directions. 
Since the expression Eq.(\ref{fourthorder}) for the fourth order term 
depends only
on the angle $\alpha_{n,m}$ between the wavevectors ${\bf q}_n$ and 
${\bf q}_m$
through $J(\alpha_{n,m})$ and $\tilde{J}(\alpha_{n,m})$, the
wavevector interaction appears therefore simply as pair interactions
depending only on the relative positions of the 2D wavevectors on the 
circle. We note
however that $\tilde{J}(\alpha_{n,m})$ corresponds to an interaction 
between two pairs of
opposite wavevectors  $({\bf q},-{\bf q})$ whereas $J(\alpha_{n,m})$ 
gives
an interaction between two single wavevectors.

We consider now the ingredients which lead to the prediction of a
cascade of transitions between order parameters with increasing number 
of wavevectors
when the temperature goes to zero. First, since $J(\alpha)$ takes very 
large positive values for
$ \alpha$ smaller than a critical angle $\alpha_0$, we can consider that 
the angle domain $ 0 < \alpha < \alpha_0$ is forbidden in order to avoid 
a dramatic increase of the free energy $\Omega$. To be quite specific 
we define $ \alpha_0$ by $J(\alpha_0) = 0$. However we could as well 
take the critical angle as $ \alpha_c$ which gives the minimum of 
$J(\alpha)$, since $ \alpha_0$ and $ \alpha_c$ are anyway very close as 
it can be seen on Fig.\ref{fig1} (we will indeed consider also $ 
\alpha_c$ in the following). On the other hand it is favorable to take $ 
\alpha $ slightly above $\alpha_0$ since it is the region where 
$J(\alpha)$ takes its most negative values. We note that the behavior of 
$\tilde{J}(\alpha)$ is not as strong compared to $J(\alpha)$, so we 
neglect $\tilde{J}(\alpha)$ in a first approximation. Next we see that 
$J(0)$ diverges as $1/T$ at low temperature and it is necessarily present 
in the free energy Eq.(\ref{fourthorder}) from the terms with ${\bf 
q}_n = {\bf q}_m$. However their unfavorable effect on $ \Omega $ is 
lowered in relative value if one increases the number $N$ of 
wavevectors, since we have $N$ terms in Eq.(\ref{fourthorder}) 
containing $J(0)$ compared to a total number of $ N ^{2}$ terms. 
Since we want to minimize $ \Omega $, this leads us to increase $N$ 
and hence to decrease the angle between wavevectors as much as 
possible. Since the angle between wavevectors is bounded from below 
by $\alpha_0$, we expect for symmetry reasons that the optimum 
structure to have regularly spaced wavevectors with an
interval angle slightly above $\alpha_0$. The last ingredient to predict 
the cascade is the fact that
$\alpha_0$ decreases with temperature. As a result, the number $N$ of 
plane waves in the optimum structure increases when the temperature 
goes to zero. This leads to a cascade of states
where the $T=0$ limit is singular. In the next sections, we present our 
arguments in more details.

\subsection{Study of the cascade}

We begin by considering only the contribution $\Omega_a$ to the free 
energy $\Omega$ from the $J(\alpha)$ terms. Even so the full problem 
of finding the order parameter structure minimizing the fourth order 
term is not a simple one. However it is natural to assume that the 
wavevectors of the $N$ plane waves have a regular angular separation 
with an angle $2 \pi /N$ between neighbouring wavevectors, so that 
their angular position is given on the circle by $ \alpha _{n} = 2 n \pi /N 
$. Otherwise we would have a minimum corresponding to a disordered 
situation for the angles, which sounds quite unlikely (note that we can 
not collapse an angle to zero since we work at fixed $N$; we will later 
on minimize with respect to $N$). Then it is easy to show that the 
weight $  w _{n} = | \Delta _{{\bf  q}_{n}}| ^{2}$ of the various 
wavevectors are all equal. Indeed the fourth order term 
Eq.(\ref{fourthorder}) is just a quadratic form in  $  w _{n}$ and 
minimizing it is formally identical, for example, to find the lowest 
energy for a single particle in a tight binding Hamiltonian on a ring, 
with hopping matrix element $ J(\alpha_{n,m})$ between site $n$ and 
site $m$ (except for the on-site term which is $ J(0)/2 $). The 
eigenvectors are plane waves and the eigenvalues are $ J(0)/2 + 
\sum_{n=1}^{N-1} J(\alpha_n) \cos(n \phi_k)$ with $ \phi_k = 2 k \pi 
/N$ and $ k = 0 , 1 , .., N - 1$. Since $ J(\alpha_n) < 0 $ for $ n \neq 0 
$, the lowest eigenvalue corresponds to $ k = 0 $ which means that the 
weight $  w _{n}$ are all equal.

Conversely if we assume from the start that the all weights $  w _{n}$ 
are equal, our problem of finding the best $ \alpha _{n}$'s is the same 
as the one of finding the equilibrium position of atoms on a ring with 
repulsive short range interaction (because $ J(\alpha)$ is large and 
positive for small $ \alpha $) and attractive long range interaction 
(because $ J(\alpha)$ is negative for larger $ \alpha $). We expect the 
equilibrium to correspond to a crystalline structure with regularly 
spaced atoms. This takes into account that $ J(\alpha)$ is a long range 
potential (clearly this regular spacing would not be the equilibrium if we 
had a strongly short range potential enforcing a specific distance 
between our atoms). Naturally in this argument we take $N$ large 
enough to fill up the ring with atoms. So we come to the conclusion that 
the minimum energy corresponds to equally spaced wavevectors.

Finally when we minimize the total free energy Eq.(\ref{freeener}) with 
respect to the weight $  w _{n}$ we find the general result:
\begin{equation}\label{minNodd}
\frac{\Omega}{N_0} = - \frac{\Omega_2 (q,\bar{\mu},T)^2}{ 
G_2(N)}
\end{equation}
where we have, in the case $ \Omega = \Omega_a$:
\begin{equation}
\label{G2odd}
N G_2(N) = 2 J(0) + 4 \sum_{n =1} ^{N-1}  J(\alpha_n)
\end{equation}
We note that we still have a degeneracy of the lowest energy 
configuration with respect
to the choice of the plane waves phases.

We now take also into account in the free energy the terms containing 
$\tilde{J}(\alpha_{n,m})$. These terms appear when there are pairs of 
opposite wavevectors
$({\bf q},-{\bf q})$. With our assumption of regularly spaced 
wavevectors, this corresponds to even $N$ - in which
case we have $N/2$ pairs - whereas for odd $N$, there are no
$\tilde{J}(\alpha_{n,m})$ terms in the free energy. Note that
$\tilde{J}(\alpha)$ is negative for any angle $\alpha$ so that
it is favorable to take pairs of opposite wavevectors. This seems to be in
favor of taking $N$ even and we will indeed see that as a result states 
with even $N$ will be selected.

For even $N$, the total free energy is:
\begin{equation}
\frac{\Omega}{N_0} = \frac{\Omega_a}{N_0} + 2 
\sum_{n=0}^{N/2-1}
\sum_{m=0}^{N/2-1}  ( 1- \delta_{n,m} )
\Delta_n \Delta_{-n} \Delta_m^* \Delta_{-m}^*
\tilde{J}(\alpha_{n-m})
\end{equation}
where we have used $\tilde{J}( \pi -\alpha) = \tilde{J} (\alpha)$
and $\alpha_{n,m} = 2 (n-m) \pi /N = \alpha_{n-m}$; $\Delta_n$ is a 
shorthand for $\Delta _{{\bf  q}_{n}}$ and so on. The fact that 
$\tilde{J}(\alpha)$ is always negative has a direct consequence on the 
phases of the plane waves. In order to minimize the free energy they 
have to be chosen so that $\Delta_n \Delta_{-n} \Delta_m^* \Delta_{-
m}^* $ is always real. Writing $\Delta_n  = |\Delta_n| e^{i 
\phi_n}$, this implies that $\phi_n + \phi_{-n} = \Phi_0 $ for any $n$, 
where $\Phi_0$ is a constant phase which can be chosen to be zero, 
since this merely corresponds to a global phase change for the order 
parameter. This link between phases for opposite wavevectors removes 
only a part of the degeneracy, since the phase $\phi_n = -\phi_{-n}$ 
can be still arbitrarily chosen. Now we see that the contribution of the 
$\tilde{J}(\alpha_{n,m})$ terms to the free energy is a quadratic form 
in $  w' _{n} =\Delta_n \Delta_{-n}$ with negative coefficients, and no 
on-site contribution. This is quite similar to what we have found for the 
$J(\alpha)$ terms in the preceding subsection. So the equal weight 
distribution, which minimizes the $J(\alpha)$ terms, gives also 
independently the minimum of the contribution of the 
$\tilde{J}(\alpha)$ terms. This means that the inclusion of these last 
terms in the free energy does not change the optimum structure
apart from the phase link between opposite wavevectors. Finally the 
total minimum free energy is still given by Eq.(\ref{minNodd}), where 
Eq.(\ref{G2odd}) is valid for odd $N$, but has to be replaced for even 
N by:
\begin{equation}
\label{G2even}
N G_2(N) = 2 J(0) + 4 \sum_{n =1} ^{N-1}  J(\alpha_n) + 4 \sum_{n 
=1} ^{N/2-1}  \tilde{J} (\alpha_n)
\end{equation}
In this last case the corresponding equilibrium order parameter is real, 
being a sum of cosines of the form $\Delta ({\bf  r}) = |\Delta_1 | \sum 
_{i} \cos({\bf  q}_{i}. {\bf  r} + \Phi_i)$.
We turn now to the minimization of the free energy with respect to 
$N$.
\begin{figure}[h]
\begin{center}
\includegraphics[width=0.6\textwidth,height=0.3
\textheight]{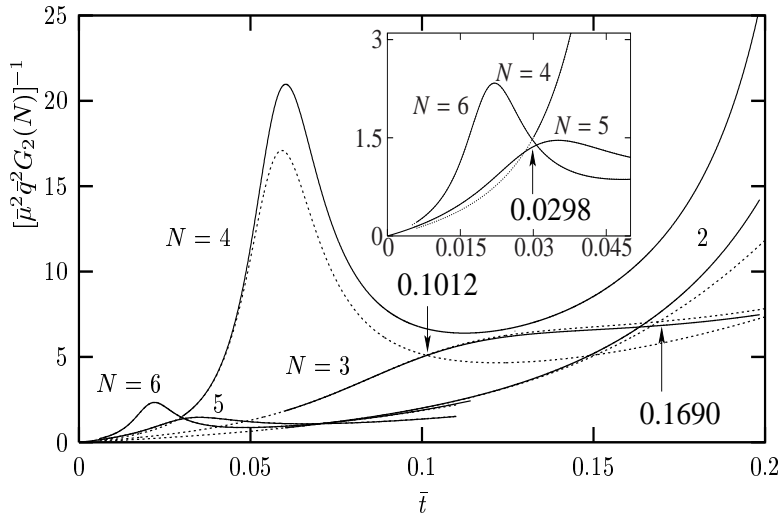}
\caption{  $[G_2 ] ^{-1}$ as a function of the reduced temperature 
$\bar{t} = T/\bar{q} \bar{\mu}$, for different values of $N$.
Full lines correspond to our analytical low temperature expressions
from Eqs. (\ref{alowtemp}), (\ref{eqJ}) and (\ref{Jtilde}), while 
dashed lines
show the exact result. They agree perfectly for $N=4,5,6$ in the
$\bar{t} \le 0.05$ domain.}
\label{fig3}
\end{center}
\end{figure}

\subsection{Minimization with respect to $N$.}

We consider now the numerical calculation of $G_2(N)$ for different 
order
parameter structures, i.e., for various values of the number of
plane waves $N$. From Eq.(\ref{minNodd}) the equilibrium order 
parameter corresponds to the maximum $[G_2 (N)] ^{-1}$. In Fig. 
\ref{fig3} we show for $[G_2 (N)] ^{-1}$ both our low temperature 
expansion and the exact
evaluation of the Matsubara sum in Eqs. (\ref{eq16}) and (\ref{eq17}),
as a function of the reduced temperature $\bar{t}$.
We see that below $\bar{t} \simeq 0.05$, the low temperature analytical
expressions, Eqs. (\ref{alowtemp}), (\ref{eqJ}) and (\ref{Jtilde}), 
agree remarkably well with
the exact result. They are therefore completely sufficient quantitatively 
to study the cascade.

Fig. \ref{fig4} displays the same quantity for lower temperatures, 
exhibiting the cascade of transitions. An interesting feature of this 
cascade is that an order
parameter with an odd number of plane waves is never the lowest 
energy
solution, except from the $N=3$ case, already found by Shimahara 
\cite{shima2} .
The reasons for this result will appear clearly in the next section, where 
we will study analytically the asymptotic regime of low temperatures.

\section{Low temperature asymptotic behaviour} \label{last}
\subsection{Minimum angle}

In the low temperature limit $t \ll 1$, we can derive an explicit
expansion for the value of the zero $ \alpha_0$ of $J(\alpha)$, as well 
as the location  $ \alpha_c$ of its minimum. These critical angles are 
crucial to study the cascade since they
give essentially the minimum angle between two wavevectors. In this 
low temperature range we have $ \bar{q} \simeq 1$, so that $ \bar{t} 
\simeq t$.
As can be seen in Fig.\ref{fig1}, $J(\alpha)$ has two extrema: a 
maximum for $ \alpha $ around $a \simeq \beta^2/2$, as it is clear from 
the $\cosh ^{-2}$ term in Eq.(\ref{Jalpha}), and, for a slightly larger 
value of $ \alpha $, a minimum at $ \alpha_c$. The zero $ \alpha_0$ is 
naturally in between. The condition $ \beta^2/2 \simeq a \simeq a_0 = 
(1/4) \ln( \pi /2t) $ implies
$\alpha \simeq 2 \sqrt{t \ln(\pi /2 t)}$. Therefore at low $t$, both $ 
\alpha_0$ and $ \alpha_c$ are in a domain where $\alpha \to 0$ and $x 
= \beta ^{2} /2 = \alpha ^{2}/16 t \to +\infty$.
In this regime the two terms in Eq.(\ref{Jalpha}) for $J(\alpha)$ 
simplify to give:
\begin{figure}[h]
\begin{center}
\includegraphics[width=0.6\textwidth,height=0.3
\textheight]{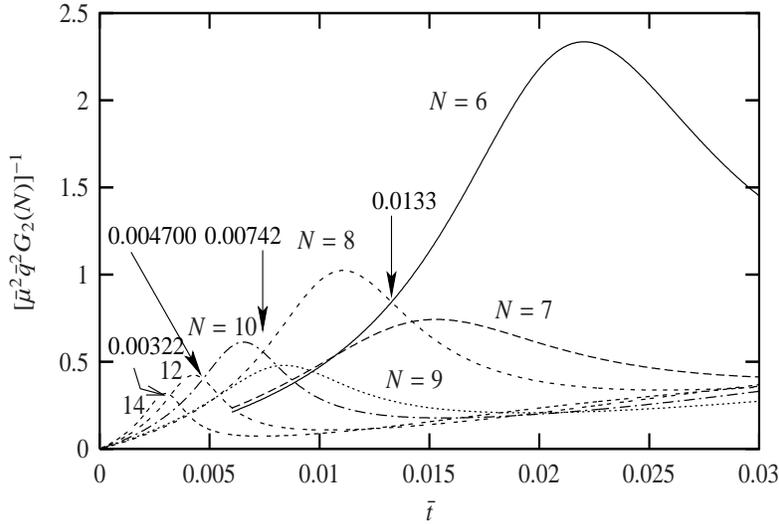}
\caption{$G_2$ plotted for temperatures lower and values of $N$ 
higher than in Fig.\ref{fig3},
using Eqs. (\ref{alowtemp}), (\ref{eqJ}) and (\ref{Jtilde}).
For clarity we do not represent solutions of $N$ larger than $14$ but
they can be easily numerically evaluated.}\label{fig4}
\end{center}
\end{figure}

\begin{equation}\label{asympto}
\bar{\mu}^2 J(\alpha) \simeq
\frac{\pi}{16 t \alpha \cosh^{2} X} - \frac{1}{\alpha^2}
\end{equation}
where $X= x - a_0$. The second term in this expression simply comes 
from $-1/4\sin^2(\alpha/2)$
which is the asymptotic limit Eq.(\ref{Jlargealpha}) for $ \bar{\mu}^2
J(\alpha)$. 

Let us first consider the calculation of $ \alpha_c$. The first term is 
responsible for the strong upturn near the minimum. From the 
derivative of Eq.(\ref{asympto}) $J(\alpha)$ is extremal for:
\begin{equation}
\cosh^{2} X = \frac{ \pi }{2} \frac{ x ^{3/2}}{t ^{1/2} } ( \tanh X + 
\frac{4}{x} )
\end{equation}
The minimum we are looking for is in the domain $ X > 0$ and from 
the above equation it is found for a large value of $X$. This allows to 
simplify this equation into:
\begin{equation}
\label{eqk}
e^{x} = k x^{3/4}
\end{equation}
with $k = (2 \pi ^{3}/ t ^{2}) ^{1/4}$. Writing this equation $ x = \ln 
k + (3/4) \ln x$ one can generate a solution by the recurrence relation $ 
x _{n+1} = \ln k + (3/4) \ln x_n$ which converges very rapidly. For 
example the second iteration gives:
\begin{equation}
x =  \ln k + (3/4) \ln [\ln k + (3/4) \ln ( \ln k )]
\label{secondit}
\end{equation}
The corresponding result for $ \alpha _c$ is to leading order in the low 
temperature limit:
\begin{equation}\label{alphac}
\alpha_c = \left( 8t \ln \frac{(2 \pi ^{3})^{1/2}}{t}\right) 
^{\frac{1}{2}}
\end{equation}
but numerically this is not such a good result at low temperature and 
one has rather to perform a few iterations to get the correct answer. For 
example in Eq.(\ref{eqk}), the exact result is $x=5.94$ for $k=100$, 
while $\ln k=4.60$, but the second iteration Eq. (\ref{secondit}) gives 
$x=5.92$.

Then in order to find the optimum number of plane waves we notice
that $J(\alpha)$ rises very rapidly below $\alpha_c$, so we can not
have basically the angular separation $ 2 \pi /N$ between two plane 
waves less
than $\alpha_c$. Since on the other hand it is energetically favorable to
take $N$ as large as possible, as long as $J(\alpha)$ is negative, we can 
find an asymptotic estimate of the optimum value $N_o$ of $N$ by 
taking the integer value of $ 2 \pi
/\alpha_c$, that is:
\begin{equation}
N_o = {\rm E}\left[ \frac{2 \pi }{\alpha_c } \right]
\label{asymptN}
\end{equation}

Following the same procedure, we can also derive an explicit 
expression
for the angle $\alpha_0$ corresponding to the zero
of $J(\alpha)$. From Eq.(\ref{asympto}), we find:
\begin{equation}
\cosh^{2} X = \frac{ \pi \alpha}{16 t }.
\end{equation}
Naturally $\alpha_0$ is close to $\alpha_c$ since $J(\alpha)$
is rapidly increasing below $\alpha_c$. We are still in the $X >0$
domain and the solution  corresponds to $X$ large. This simplifies the 
above equation
into:
\begin{equation}
e^{x} = \frac{k}{\sqrt{2}} x^{1/4}
\label{eqkbis}
\end{equation}
where $k$ has been defined previously.  The corresponding  recurrence 
relation
which generates the exact solution is
$ x _{n+1} = \ln (k/\sqrt{2}) + (1/4) \ln x_n$. The second iteration 
gives
here:
\begin{equation}
x =  \ln  \frac{k}{\sqrt{2}}
 + \frac 1 4 \ln \left[
\ln \frac{k}{\sqrt{2}}  + \frac 1 4
\ln \left( \ln \frac{k}{\sqrt{2}}  \right) \right]
\end{equation}
and $\alpha_0$ is simply given by $\alpha_0 = \sqrt{16 t x}$. We can 
then make the same argument as above for $ \alpha_c$, and write the 
following asymptotic estimate of the optimum value $N_o$:
\begin{equation}
N_o = {\rm E}\left[ \frac{2 \pi }{\alpha_0 } \right]
\end{equation}
which will naturally be found to be quite near the above one 
Eq.(\ref{asymptN}), since  $\alpha_0$ and  $\alpha_c$ are quite close.

\begin{figure}[h]
\begin{center}
\includegraphics[width=0.6\textwidth,height=0.3
\textheight]{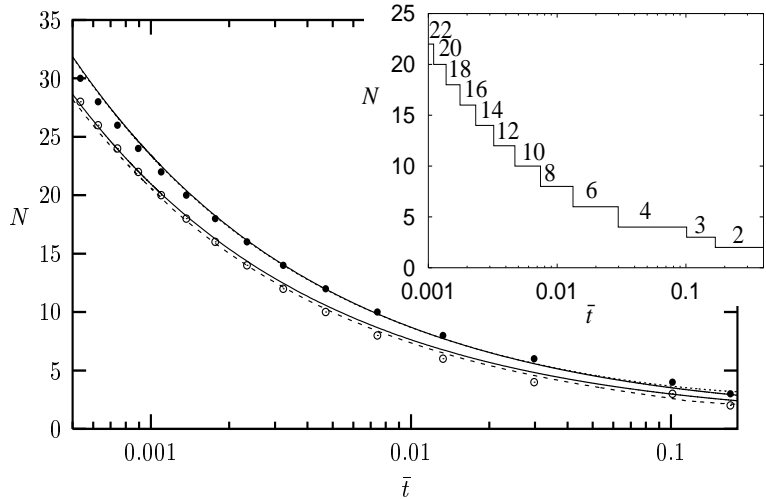}
\caption{ Optimal value of $N$ and locations of the critical
temperatures. The dots give the temperatures where
the optimal $N$ changes: the white dot is the optimal
$N$ above the critical temperature whereas the black dot
is the optimal $N$ below. $2 \pi/\alpha_c$ (curves very near the white 
dots)
and $2 \pi / \alpha_0$ (curves very near the black dots) are
also shown on this figure. The dashed lines show the exact
calculations,while the solid lines are determined with the recurrence
formulas below Eq. (\ref{eqk}) and Eq. (\ref{eqkbis}) with only one 
iteration.
In the insert for clarity we only show the "devil" staircase 
corresponding to the white and black dots.
 }\label{figcompare}
\end{center}
\end{figure}

Naturally we can calculate numerically exactly the optimal value $N_0$ 
of $N$ as a function
of the temperature, from our above results. In the same process we find 
also the
critical temperatures where the $N_0$ changes. They are given by
the crossings of the $G_2(N)$ curves for different values of $N$, as it 
is seen on Fig. \ref{fig3} and \ref{fig4}.
In Fig. \ref{figcompare} these exact critical temperatures are compared
with the exact calculations of $2 \pi/\alpha_c$ and $2 \pi / \alpha_0$, as 
well as their asymptotic values. As it is seen on this figure, it happens 
that the optimal values of $N$ falls essentially just between $2 
\pi/\alpha_c$ and $2 \pi / \alpha_0$ in the low temperature domain.

\subsection{Asymptotic evaluation of $G_2 (N)$}

It can be seen on Fig. \ref{fig1} that $J(\alpha)$ switches rapidly above 
$\alpha_c $ to its large angle asymptotic behaviour:
\begin{equation}
 \bar{\mu}^2  J (\alpha) \simeq -\frac{1}{4 \sin^2(\alpha/2)}
\label{asymptotic}
\end{equation}
so we may with a fair precision use this simplified expression for $ J 
(\alpha_n)$, with $\alpha_n = 2 \pi n/N$, to evaluate the sum in 
Eq.(\ref{G2odd}) for $G_2 (N)$. At low temperature the number of 
plane waves N is large and the sum is dominated by the terms with 
small $n$ corresponding to small angles, for which we have $\sin 
(\alpha/2) \simeq \alpha /2$. Hence we have for the sum coming in Eq. 
(\ref{G2odd}) for $G_2 (N)$:
\begin{equation}
\frac{4}{N} \sum_{n=1}^{N-1} \bar{\mu}^2 J (\frac{2 \pi n}{N})
\simeq - \frac{8}{N} \sum_{n=1}^{N/2-1}  \frac{1}{4 (n \pi/N) ^2}
= -\frac{N}{3} + {\mathcal O} \left(\frac{1}{N}\right)
\end{equation}
For even $N$, we have also to evaluate  the contribution  coming from 
the  terms in Eq. (\ref{G2even})
containing $\tilde{J}( \alpha )$. Again we can use for most angles 
the asymptotic expression Eq.(\ref{jtildelim}) for $\tilde{J} (2 \pi n 
/N)$. Once more the dominant contribution comes from the small $n$ 
terms, and the logarithmic leading behaviour is accurately given by the 
asymptotic expression of $\tilde{J}( \alpha )$ as:
\begin{equation}
\frac{4}{N} \sum_{n=1}^{N/2-1} \bar{\mu}^2 \tilde{J}
 ( \frac{2 \pi n}{N}) \simeq - \frac{2 \pi}{N} \sum_{n=1}^{N/4}
\frac{1}{2 \pi n /N} = - \ln N + {\mathcal O} (1)
\end{equation}
[VERIFIER le n=1]We see here explicitely that $J(\alpha)$ has a 
dominant role compared to
$\tilde{J}(\alpha)$.
Taking into account the $J(0)$ contributions, the dominant behaviour of 
$G_2 (N)$  can be summarized in the limit of
 large plane wave number $N$ and low  temperature as:
\begin{eqnarray}\label{G21}
\bar{\mu}^2 G_2(N) &\simeq& \frac{1}{2 N t} - \frac{N}{3}
 \qquad \qquad \qquad \qquad \qquad \qquad N \textrm{odd} \\
\label{G22}
\bar{\mu}^2 G_2(N) &\simeq& \frac{1}{2 N t} - \frac{N}{3}
- \ln N + cte
 \qquad \qquad \qquad  N \textrm{even}
\end{eqnarray}
These expressions correspond actually in Fig. \ref{fig4} only to the 
rising part of $ [G_2(N)] ^{-1}$ on the low temperature side. The 
downturn of $ [G_2(N)] ^{-1}$ for higher temperature is due to 
contributions from the positive part of $ J (\alpha)$ which are beyond 
our asymptotic approximation Eq. (\ref{asymptotic}). Nevertheless we 
have also found above the critical temperatures for switching from a 
given value of $N$ to the next one. From Fig. \ref{fig4} it is seen that 
it is enough to plug these critical temperatures into the evaluation of the 
rising part of $ [G_2(N)] ^{-1}$ to obtain the free energy at the 
transitions. When we substitute accordingly in these expressions Eq. 
(\ref{G21}) or (\ref{G22}) the value $N = 2 \pi / \alpha _c$ for the 
optimum plane wave number we have: 
\begin{equation}
\bar{\mu}^2 G_2(N) = \frac{\alpha_c}{4 \pi t} \left( 1-
\frac{8 \pi^2 t}{3 \alpha_c^2} \right) + {\mathcal O} (\ln(1/t))
\end{equation}
Since $ 8t / \alpha_c^2 \sim 1 / \ln(1/t)$ from Eq.(\ref{alphac}), we 
find that $ G_2(N)$ is always positive in the low temperature range 
(more precisely we have $ 1- (8 \pi^2 t)/(3 \alpha_c^2) =1- \pi 
^{2}/6x$ which is positive as soon as $x > \pi ^{2}/6$, i.e. from 
Eq.(\ref{eqk}) exactly for $t<0.62$). This means that the transition 
stays always second order (a negative sign would have implied a first 
order transition and a breakdown of our fourth order expansion).

Finally it seems from Eq.(\ref{G21}-\ref{G22}) that it is favourable to 
have an even number of plane waves in order to take advantage of the 
additional $\tilde{J}$ contribution. This can be confirmed by a more 
careful comparison between two consecutives values of $N$. Let us 
assume that $N$ is odd with $N = 2 \pi / \alpha _c$ and compare $ 
G_2(N)$ to $ G_2(N-1)$. By going from $N$ to $N-1$ we gain 
naturally the $\tilde{J}$ term in Eq.(\ref{G22}), but we increase the 
first term due to the unfavorable of the $J(0)$ term. Nevertheless:
\begin{equation}
\bar{\mu}^2 [G_2(N-1) - G_2(N)] = \frac{1}{2 N(N-1) t} - \ln(N-1) 
+ {\mathcal O}(1).
\end{equation}
where the dominant term becomes exactly with the help of 
Eq.(\ref{eqk}):
\begin{equation}
\bar{\mu}^2 [G_2(N-1) - G_2(N)] = - x ( 1 - \frac{2}{\pi^2}) + 
\frac{5}{4} \ln x + \frac{1}{4} \ln \frac{32}{ \pi }
\end{equation}
which is always negative since its maximum, reached for $ 1/x_m = 
(4/5) (1-2/ \pi ^{2})$, is -0.107.
Therefore the asymptotic evaluation of $G_2(N)$ shows indeed 
explicitely that the only order parameters which appears at low 
temperature are those with an even number of plane waves, in 
agreement with our exact numerical results.

\subsection{Phenomenological interpretation}

It is interesting to compare our results
to the pairing ring picture explored by Bowers and Rajagopal 
\cite{bowers} 
(BR) in the three dimensional case.
BR have looked at the free energy expansion at zero temperature
in $3$D, extending the work of Larkin and Ovchinnikov.
In the case of multiple plane waves in the order parameter,
they have pointed out a simple physical interpretation of their results.
Their picture is based on the Fulde and Ferrell study \cite{ff} of a single 
plane
wave order parameter where they showed that one may separate the 
wavevector space
in two complementary domains: the pairing region and the pair breaking
region. In the case of a vanishing order parameter and considering only 
wavevectors at the Fermi surface, the pairing region is merely an 
infinitely thin circle. It is given
by the intersection of the up and down spin Fermi surfaces, after 
shifting one of them by $ {\bf  q}$.  
The total opening angle of this circle is thereforer given by $\psi_0  = 
\arccos(1/\bar{q})$. In
$3$D and at $T=0$, $\bar{q}=1.1997$ which gives $\psi_0  = 
67.1^{\circ}$.
For a non-zero order parameter amplitude $\Delta$, the circle broadens 
and
becomes a ring whose width is given by $\Delta/\bar{\mu}$. 
Therefore, the pairing region for a single plane wave is
a ring drawn on the Fermi surface, whose center is on
the plane wave direction ${\bf  q}$ (the restriction to the Fermi surface 
is justified by the fact that only wavevectors close to the Fermi surface 
are relevant for pairing).

In the case of multiple plane waves in the order parameter,
BR showed that the intersection of pairing circles
from different plane waves is energetically largely unfavored,
the worst case being when two circles are exactly in contact.
The maximisation of the pairing regions leads naturally to increase as 
much as possible the plane waves number $N$ with the constraint
that they are no intersecting circles. This leads to nine circles
on the Fermi surface. BR rather concluded that
the least energy state was the one with eight
plane waves for 'regularity' reasons. In particular,
this leads to a real order parameter. We see now that
our cascade in $2$D is a direct consequence
of these two heuristic principles : the no-intersecting circles
rule and the maximisation of plane waves.

In $2$D, the circles are replaced by pairs of points
separated by the angle $\psi_0$, and the no-intersecting rule becomes 
now a no-entanglement rule between different pairs. But this is just 
equivalent to our above finding from Fig. \ref{fig1} that there is an 
effective short range repulsion between different wavevectors ${\bf  
q}$, with indeed the worst case corresponding to the contact since the 
maximum of $J( \alpha )$ is just below its zero. The contact condition 
$a=\bar{\beta}^2/2$ that we have found (corresponding to the 
maximum of $J( \alpha )$) can be rewritten as $\cos(\alpha/2) = 
1/\bar{q}$ which implies $\alpha /2 = \psi_0 /2 $. This is exactly the 
same as the contact condition of BR. Next we have found explicitely 
that it is indeed favorable to maximize the number of plane waves, 
taking the short range repulsion into account. Our cascade is due to the 
fact that, when the temperature
goes to zero, $\psi_0$ also drops to zero and the pairing region  for a 
single plane
wave shrinks to a single point, corresponding to the situation where the 
two up and down Fermi circles are just in contact.

\section{ CONCLUSION }

In this paper we have investigated the low temperature range for the 
FFLO transition in two dimensions and we have shown that the order 
parameter is no longer a simple $ \cos({\bf  q}. {\bf  r})$ at the second 
order transition, in contrast with the situation found near the tricritical 
point. Instead the transition is toward more and more complex order 
parameters when the temperature goes to zero, which gives rise to a 
cascade with, in principle, an infinite number of transitions. At the 
transition these order parameters are in general a superposition of an 
even number of plane waves with equal weight and equal wavevector 
modulus, which corresponds to a real order parameter equal to a 
superposition of cosines. The directions of these wavevectors are found 
to be equally spaced angularly, with a spacing which goes to zero when 
the temperature goes to zero, which is the reason for the ever increasing 
number of plane waves in this limit. The singular behaviour in this limit 
$ T = 0$ is actually present in all physical quantities. It arises because, 
in order to obtain the lowest energy, the two Fermi circles 
corresponding to opposite spins come just in contact when one applies 
the shift corresponding to the wavevector $ {\bf  q}$ of the FFLO 
phase. This is this situation, with these two circles just touching each 
other, which gives rise to the singularities. Naturally this is linked to the 
fact that in 2D the Fermi surface is actually a line. Hence this singularity 
is a general feature of 2D physics and we may expect it to give rise to 
similar consequences in more realistic and more complex models 
describing actual physical systems.

Naturally in this paper we have addressed this remarkable situation in 
the most simple physical frame. We have considered the simplest BCS 
model with isotropic Fermi surface, that is a circle for our 2D case. 
Naturally if anisotropy is introduced in the dispersion relation of the 
electrons, or in the effective electron-electron interaction, or both, the 
physics will be more complex. In the same way we expect Fermi liquid 
effects \cite{br} to bring important quantitative modifications. Similarly 
if we consider a quasi 2D superconductor with a magnetic field not 
perfectly aligned with the planes, currents will be produced and orbital 
contributions to the free energy will arise. However in order to have the 
physics right in all these complex situations, it is quite clear that one has 
to obtain the correct limit in the simplest possible case that we have 
considered in the present paper. 

\section{Appendix A}

In this section we rederive rapidly our result Eq.(\ref{leading2}) for the 
leading order of the optimal wavevector at low temperature, taking the 
same starting point as Bulaevskii \cite{bul}.

We start from Eq.11 of Ref.\cite{bul}:
\begin{eqnarray}
\ln(T_c/T) =  \frac{1}{\pi}{\rm Re} \int_{- P v_F} ^{ P v_F}
d \Omega  \frac{1}{\sqrt{P^2 v_F^2 - \Omega ^2}}
[ \psi  (1/2 + i (2 \mu_{0} H + \Omega)/4 \pi T)) - \psi  (1/2)]
\end{eqnarray}
We go to our notations by setting $ H = \bar{ \mu }, P = q $ and
introduce as above our reduced variables $ \bar{q} = q v_{F}/2 
\bar{\mu }$
and $ t = T /  \bar{\mu}$. When we perform the angular integration as
in Ref. \cite{bul} by setting $\Omega = P v_F \cos \theta $ and 
introduce for the
digamma function the integral representation (Eq.13 of Ref. \cite{bul}):
\begin{eqnarray}
\psi  (x) = \ln x - 1/2x - 2 \int_{0} ^{ \infty  }
\frac{y dy}{(y^2 + x^2) ( \exp (2 \pi y) - 1)}
\end{eqnarray}
we obtain Eq.14 of Ref. \cite{bul} (correcting some minor misprints). 
We can then
expand the result to all orders in $t$ and show that all the terms in the
expansion are zero. However rather than displaying the steps of this
calculation here, it is more convenient to immediately remark that the
optimal $ \bar{q}$ is obtained by writing that the derivative of Eq.11
with respect to $ \bar{q}$ is zero. This provides an equivalent 
calculation
(we work on the derivative instead of working on the function), which 
is
somewhat easier and allows us also to make in the following the direct 
contact
with our result. This condition on the derivative, which gives the 
equation
for the optimal $ \bar{q}$, is:
\begin{eqnarray}
{\rm Im} \int_{0} ^{ \pi } \frac{d \theta} {\pi}
\psi' (1/2 + i (1 + \bar{q} \cos \theta )/2 \pi t)) = 0
\end{eqnarray}

Taking the derivative of the above integral representation, we obtain a
corresponding representation for $\psi'$ on which the angular 
integration
is easily performed (the integrals can be found in Gradstein and Ryzhik 
\cite{gr}).
We display the result by introducing the function $ f(x) = x / \sqrt(x^2-
1)$
and setting $ x_0 = 1/\bar{q}$ , $ \epsilon = - i \pi t /\bar{q}$ and
$ x = x_0 + \epsilon $. The result is:
\begin{eqnarray}
{\rm Re} \, [ 1 - f(x) + \epsilon  f'(x) - \frac{\pi}{2} \int_{-\infty} ^{ 
\infty  }
\frac{dy}{\sinh^2 ( \pi y) } ( f(x) - f(x_0 + \epsilon (1+2iy)) ) ] = 0
\label{condition}
\end{eqnarray}
where the integral goes from $ - \infty $ to $ \infty $ because we have
collected two terms into one. Now we can write for the first three terms
the Taylor expansion:
\begin{eqnarray}
1 - f(x) + \epsilon  f'(x) = 1 - f(x_0) + \sum_{0}^{\infty }
\frac{\epsilon ^{n+1}}{n!} \frac{n}{n+1} f ^{(n+1)} (x_0)
\end{eqnarray}
When we take the real part, only the odd order derivatives of $f$
contribute because $ \bar{q}$ is near $1$ with $ \bar{q} > 1$, which
makes $ x_0 < 1$ and $ Re f ^{(2p)} (x_0) = 0$. This gives:
\begin{eqnarray}
{\rm Re}\,  [ 1 - f(x) + \epsilon  f'(x) ] = 1 + \sum_{p=1}^{\infty }
\frac{\epsilon ^{2p+1}}{(2p)!} \frac{2p}{2p+1} f ^{(2p+1)} (x_0)
\end{eqnarray}
Similarly we can perform the expansion in the integral and expand
$ (1+2iy) ^n$ by introducing the binomial coefficients $ C _{n}^{p}$.
All the resulting integrals can be found in Gradstein and Ryzhik 
\cite{gr},
and expressed in terms of Bernoulli numbers $B_{2m}$. This leads to:
\begin{eqnarray}
\frac{\pi}{2} {\rm Re} \,  \int_{-\infty} ^{ \infty  }
\frac{dy}{\sinh^2 ( \pi y ) } [ f(x) - f(x_0 + \epsilon (1+2iy)) ] =
 \sum_{p=1}^{\infty } \frac{\epsilon ^{2p+1}}{(2p+1)!} f ^{(2p+1)} 
(x_0)
\sum_{m=1}^{p}  C _{2p+1}^{2m} 4^m B_{2m}
\end{eqnarray}
Now a remarkable identity (found for example in Gradstein and Ryzhik 
\cite{gr})
for Bernoulli numbers states that:
\begin{eqnarray}
\sum_{m=1}^{p}  C _{2p+1}^{2m} 4^m B_{2m} = 2p
\end{eqnarray}
As a result, by gathering all the terms, all the coefficients of the
powers of $ \epsilon $ are zero, and it is not possible to satisfy the
condition that the above derivative is zero. The answer to this puzzle
is that the contribution of the terms we have considered is not exactly
zero, but exponentially small, which explains why we find it to be
zero in a perturbative expansion. This is shown now in the
following.

We start again from the above condition Eq.(\ref{condition}) found for 
the optimum
wavevector. We transform the integral by shifting the integration 
contour toward
the upper complex plane by $ i/2$ for the variable $y$. First we have
to take care that integrant in the above integral has no singularity
for $y=0$, because proper cancellation between various terms. Hence
we can also say that this integral is equal to its principal part.
Next if we want a complete contour $C$, we have to add to this 
principal
part the contribution of an infinitesimal semi-circle around
$ y=0$ with positive imaginary part. This contribution is easily
found by residues to be equal to $ - \epsilon f'(x)$, so we have:
\begin{eqnarray}
{\rm Re} \, [ \epsilon  f'(x) - \frac{\pi}{2} \int_{-\infty} ^{ \infty  }
\frac{dy}{\sinh^2 ( \pi y) } ( f(x) - f(x_0 + \epsilon (1+2iy)) ) ] =
- \frac{\pi}{2} {\rm Re}\,  \int_{C}
\frac{dy}{\sinh^2 (\pi y)} [ f(x) - f(x_0 + \epsilon (1+2iy)) ]
\end{eqnarray}
Now we set $ \pi y = z + i \pi /2$ and make use of $ i\sinh(z+i \pi /2)
= i \cosh(z)$ which gives:
\begin{eqnarray}
{\rm Re} \, [ \epsilon  f'(x) - \frac{\pi}{2} \int_{-\infty} ^{ \infty  }
\frac{dy}{\sinh^2 ( \pi y )} ( f(x) - f(x_0 + \epsilon (1+2iy)) ) ] =
\frac{1}{2} {\rm Re}\, \int_{-\infty} ^{ \infty  }
\frac{dy}{\cosh^2 z} ( f(x) - f(x_0 + 2tz/\bar{q} ) )
\end{eqnarray}
where the contour runs infinitesimally below the real $z$ axis.
The first term in the last integral is just equal to $f(x)$, so our
equation becomes:
\begin{eqnarray}
{\rm Re}\, [ 1 - \frac{1}{2} \int_{-\infty} ^{ \infty  }
\frac{dy}{\cosh^2 (\pi y)} f(x_0 + 2tz/\bar{q} )] = 0
\end{eqnarray}
It is easily checked that this equation is identical to our 
Eq.(\ref{leading}).
Note that the fact that
the contour runs infinitesimally above or below the real axis is
unimportant since we take the real part, and the contribution along the
cut due to the square root is purely imaginary.

In conclusion we have obtained our basic equation for the second order 
term by taking the same
starting point as Ref. \cite{bul}.  The end of the argument to obtain 
Eq.(\ref{leading2})
is naturally the same as following our Eq.(\ref{leading}).
In particular this argument shows that the integral in the above formula
is proportional to $ \exp{[-(\bar{q}-1)/t]}$, so it can not be expanded
in powers of $t$.

\vspace{4mm} 
* Laboratoire associ\'e au Centre National
de la Recherche Scientifique et aux Universit\'es Paris 6 et Paris 7.


\begin{references}
\bibitem{remark}More precisely this is only valid to lowest order in 
$t$. To next order the divergence in $ 1/ \alpha $ in the two terms in 
Eq.(\ref{Jalpha}) do not cancel exactly, which gives rise to the 
divergent behaviour seen in Fig. \ref{fig1} for the dashed-dotted curve (and 
a similar behaviour occurs for $\tilde{J}$). Our temperature corrections 
eliminate this next order problem.
\end{references}
\end{document}